# Survey of Network Intrusion Detection Methods from the Perspective of the Knowledge Discovery in Databases Process

**Borja Molina-Coronado**[*]
Intelligent Systems Group
Dept. of Computer Architecture and Technology
University of the Basque Country (UPV-EHU)
Donostia (Spain)
`borja.molina@ehu.eus`

**Usue Mori**
Intelligent Systems Group
Dept. of Computer Science and Artificial Intelligence
University of the Basque Country (UPV-EHU)
Leioa (Spain)
`usue.mori@ehu.eus`

**Alexander Mendiburu**
Intelligent Systems Group
Dept. of Computer Architecture and Technology
University of the Basque Country (UPV-EHU)
Donostia (Spain)
`alexander.mendiburu@ehu.eus`

**Jose Miguel-Alonso**
Intelligent Systems Group
Dept. of Computer Architecture and Technology
University of the Basque Country (UPV-EHU)
Donostia (Spain)
`j.miguel@ehu.eus`

## Abstract

The identification of cyberattacks which target information and communication systems has been a focus of the research community for years. Network intrusion detection is a complex problem which presents a diverse number of challenges. Many attacks currently remain undetected, while newer ones emerge due to the proliferation of connected devices and the evolution of communication technology. In this survey, we review the methods that have been applied to network data with the purpose of developing an intrusion detector, but contrary to previous reviews in the area, we analyze them from the perspective of the Knowledge Discovery in Databases (KDD) process. As such, we discuss the techniques used for the capture, preparation and transformation of the data, as well as, the data mining and evaluation methods. In addition, we also present the characteristics and motivations behind the use of each of these techniques and propose more adequate

---

[*]Corresponding author

| Survey | Data selection | Data preprocessing and transformation | Data mining | Model update |
|---|---|---|---|---|
| Garcia et al., (2009) [7] | | | ✓* | |
| Sperotto et al., (2010) [8] | | | ✓* | |
| Davis et al., (2011) [9] | ✓ | ✓* | | |
| Bhuyan et al., (2014) [10] | | | ✓* | |
| Ahmed et al., (2015) [11] | | | ✓* | |
| Buzcak et al., (2016) [12] | | | ✓ | |
| Umer et al., (2017) [13] | | | ✓* | |
| Moustafa et al., (2019) [14] | ✓* | ✓* | ✓* | |

*Partial coverage

Table 1: List of published NIDS surveys, along with their coverage of methods to perform different knowledge discovery tasks.

and up-to-date taxonomies and definitions for intrusion detectors based on the terminology used in the area of data mining and KDD. Special importance is given to the evaluation procedures followed to assess the different detectors, discussing their applicability in current real networks. Finally, as a result of this literature review, we investigate some open issues which will need to be considered for further research in the area of network security.

# 1 Introduction

Intrusion Detection Systems (IDSs) are deployed to uncover cyberattacks that may harm information systems. An IDS works by analyzing different kinds of data: operating system-related data (Host-based Intrusion Detection System, HIDS) or network-related data (Network Intrusion Detection System, NIDS) [1].

HIDSs are designed to detect cyberattacks targeting an individual computer, typically a server. To do so, they process local host data such as operating system calls made by applications, log information or network traffic that reaches the machine [2]. NIDSs operate differently, focusing exclusively on network traffic, with the aim of finding malicious patterns targeting all the machines or devices inside the protected network [3]. Throughout this paper, we will pay attention to this class of IDSs, and the terms "IDS" and "NIDS" will be used interchangeably.

Finding attack indicators in network traffic data presents important particularities [4–6] such as (1) the growing number of attacks, which can evolve and change when new vulnerabilities are identified ; (2) the diverse and evolving nature of network traffic; (3) the continuously growing data volume, which challenges the successful analysis of network traffic; and (4) the increasingly common use of encrypted data, not accessible to detectors. These facts, together with the need for responses in near real-time, make intrusion detection a very hard problem.

Data mining techniques, which are applied successfully in many fields, have also proven useful to implement NIDSs. These methods can find complex relationships in the data, which can be used to detect cyberattacks. However, off-the-shelf data mining algorithms cannot be applied directly to network data, as they cannot cope with the particularities listed above. Intrusion detection is the result of a complex process that starts with the capture of network data and continues with the preparation and preprocessing of this data. Only then may a detector be able to deliver meaningful results.

In spite of this, NIDS research normally focuses only on the application of data mining algorithms to network datasets, paying minimum attention (or no attention at all) to questions such as: Which type of data (raw traffic, flows, encrypted/plaintext/no payload, etc.) will the NIDS receive? Can all the data traversing the network be gathered and processed, or just an excerpt of it? Where will the NIDS be located? How good is the performance of the NIDS in terms of both accuracy and response time? Can the detection model be updated easily to detect new threats? In this paper we aim to provide some answers to these questions.

Several surveys of NIDS research have been published already. Their contents are summarized in Table 1. As can be seen, they focus mainly on analyzing detection methods, i.e., the data mining algorithms used to identify threats, ignoring data acquisition and curation methods as well as the procedures to update the detectors' underlying models. In many cases, they are restricted in scope. For example, some only review a subset of the existing detection techniques related to the discovery of anomalous network traffic [7, 10, 11, 14, 15]; other surveys [8, 13] focus only on approaches that use a specific type of input data and ignore detectors that use other data types. These surveys have been useful to summarize and discuss NIDS proposals, but we consider that a broader vision is needed in order to understand all the building blocks that constitute an operational NIDS. This is what we aim to provide with this survey.

In this work we review NIDS proposals using the Knowledge Discovery in Databases (KDD) process as the baseline. KDD describes the procedures commonly used to find explainable patterns in data, allowing the interpretation or prediction of future events [16], and the data mining phase is only a stage of this process. Thus, like previous published surveys in the area (see Table 1), we provide an up-to-date review of the most cited and influential NIDS articles, but we also:

- Identify the techniques applied to obtain, preprocess and transform the data, discussing their motivation, assumptions and possible limitations.
- Provide an analysis of NIDS data mining methods (as other surveys do), but using two different criteria: (a) the detection approach and (b) the learning approach. Both determine how a detector behaves and are important to examine if NIDS methods are suitable for real world networks. The data mining terminology allows some terms used in misleading ways in other surveys, such as misuse and anomaly-based detection, to be clarified, avoiding the miscategorization of some NIDS proposals. We also review mechanisms to avoid deterioration in detection ability due to the evolution of network traffic, including attacks [17].
- Investigate NIDS validation and evaluation procedures. Most NIDS proposals have been evaluated considering only their detection accuracy over old-fashioned databases, something that is of very limited usefulness when considering current attacks.
- Discuss present challenges on NIDS research based on the study of the surveyed NIDS literature. We analyze the overall process and steps followed by authors to build NIDSs, identifying common assumptions and mistakes, and highlighting future research lines in the area of network security.

The remainder of this paper is organized as follows. First, we briefly depict some basic concepts about how cyberattacks can manifest themselves and describe the KDD process and its phases. In the following sections we enumerate and categorize the most representative techniques proposed in the intrusion detection literature for the different phases of the KDD process. We start with data selection and the extraction of traffic data. In Sections 4 and 5, we analyze methods for the extraction of structured feature records from that data, as well as the transformations that are commonly applied over those features. Techniques to remove data redundancies and accelerate the application of data mining algorithms are discussed in Section 6. In Section 7 we pay attention to the data mining phase, analyzing different learning methods and detection approaches. In Section 8 we show the NIDS evaluation criteria used by researchers, and also discuss the applicability of the reviewed proposals to real world scenarios. This panoramic view of NIDS proposals allows us to close the survey in Section 9 discussing present and future challenges in NIDS research, identifying common mistakes and unrealistic assumptions in the network security area.

## 2 Basic Concepts

In this section we introduce a collection of basic concepts that are necessary to describe how a NIDS operates and to understand the rest of this paper. First, we focus on cyberattacks (those that a NIDS must detect) and the way they manifest if network traffic is analyzed properly. We also describe the Knowledge Discovery in Databases (KDD) process and its different phases.

### 2.1 Networks Under Attack

Cyberattacks can be characterized by the information security properties they compromise, which are known as the CIA triad: confidentiality, integrity and availability. Confidentiality refers to secrecy of data, integrity to non-alteration of data by third parties and availability to the readiness to provide the intended service. An action is considered a network attack if it negatively affects at least one of these properties, and its attack vector (media) is the network. Thus, they should be detected through an analysis of network traffic. Table 2 lists some broad groups of known cyberattacks.

| Attack | Confidentiality | Integrity | Availability | Description |
|---|---|---|---|---|
| **Brute force** | ✓ | | | Repetitive attempt to carry out some action, such as authentication or resource discovery, often guided by a dictionary. |
| **Injection (SQL, command...)** | ✓ | ✓ | | Exploitation of weaknesses in input field sanitizing mechanisms to execute code or commands on the target system. |
| **Privilege escalation** | ✓ | | | Attempt to gain unauthorized access from an unprivileged account. |
| **Spoofing** | ✓ | ✓ | | A person or program masquerades as another, falsifying data to gain an illegitimate advantage. |
| **Sniffing** | ✓ | | | Eavesdropping traffic in order to obtain information. |
| **(Distributed) denial of service** | | | ✓ | Resource abuse intended to deny access to a service to its legitimate users. |

Table 2: A selection of generic network attacks, categorized by the system property they compromise.

To perform intrusion detection using network data, undesirable behaviors (attacks) need to be visible or latent in the data collected by the NIDS. Cyberattacks present very different purposes and characteristics and their symptoms are not always obvious. They may be visible at different vantage points and may appear at different traffic levels [18]. Attack manifestations can be classified as:

- **Point**. The malicious activity affects one data sample, e.g., a packet, a connection. In this case, an attack leaves a footprint in a single record, meaning that some of their values are not permitted or constitute an anomaly. For example, SQL injection is carried out by placing code in the parameters of a HTTP request with the aim of manipulating the behavior of a web application. This attack can be detected by analyzing a single sample of HTTP payload.
- **Contextual**. A particular event is dangerous but only in the specific situation in which it occurs. The event itself is not an anomaly, but may indicate that an attack is happening due to its unexpected appearance given the context. For example, a HTTP (web) request to a DNS server is not an anomaly by itself, but could be indicative of a service discovery attempt over a server.
- **Collective**. The attack shows up as several characteristic events which, together but not individually, constitute a malicious event. For example, a large number of simultaneous requests to a common service form an anomaly, although each individual connection is not anomalous by itself.

Given the diverse forms in which a malicious event may leave a footprint, different strategies need to be taken during the construction of a NIDS to allow detection in a satisfactory way. Different views of network traffic are required to recognize different cyberattacks, and successful identification is not only related to the choice of a good data mining algorithm, but also to the information available (captured and processed) to feed it.

## 2.2 The Knowledge Discovery in Databases Process

The Knowledge Discovery in Databases (KDD) process describes the procedures commonly used to find explainable patterns in data, allowing the interpretation or prediction of future events [16]. KDD was defined formally in [19] as:

> *"The nontrivial process of identifying valid, novel, potentially useful, and ultimately understandable patterns in data."*

KDD is an iterative process, as depicted in Figure 1, comprising the following steps [16]:

- **Data collection** consists of selecting and capturing raw data from their sources. The data gathered in this step should provide the necessary information to solve the problem at hand.
- **Data preprocessing** involves the extraction of a valid and structured set of those features required for knowledge extraction. It also includes the application of cleansing methods to avoid factors such as noise, outliers or missing values, and the derivation of new features from others to structure the data in a way appropriate for the application of the algorithms of the following phases.

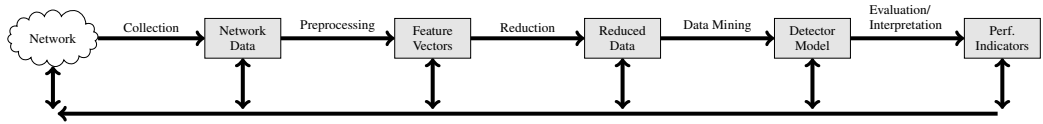

Figure 1: KDD flow for IDS, showing the results at each step and its iterative nature.

- **Data reduction** is a necessary step to optimize the application of data mining algorithms, reducing their computation time and/or improving their results. It involves feature subset selection (removal of useless or redundant features), dimensionality reduction (projection of a set of features into a smaller space) or modifications of the set of samples (adding or removing samples).
- **Data mining** is the core of the KDD process, but it can be successful only if the previous steps are performed properly. It consists of selecting and applying techniques, including machine learning algorithms, to extract knowledge from data by finding useful patterns and relations between features. The choice of a specific data mining algorithm depends on the problem at hand and the nature of the available data.
- **Interpretation and evaluation** consists of measuring the performance and effectiveness of the data mining algorithm using a set of metrics, and, consequently, of the tasks carried out in previous steps of the KDD process. It also comprises result visualization and reporting.

Given the iterative nature of this process, note that some steps may have to be reapplied and rethought more than once to achieve satisfactory results [20]. For example, a feature initially excluded during the data transformation step may be added later in order to increase the performance of a particular data mining algorithm.

## 3 Data Collection Stage

The first phase of the KDD process involves the collection of data. Network activity data can be gathered at different locations. Network devices may provide two different types of data: raw packets and flow records.

Commonly, networks are organized in a hierarchical (tree) structure where the devices (typically, routers and switches) have visibility of the traffic entering and leaving the machines connected below them. Higher level (root or core) devices have a complete view of Internet-related traffic going to and coming from the computers (and network devices) located below their level. However, they have a limited view of horizontal communications (those between machines at the same level). The opposite is true for lower level (aggregation and access) devices. Thus, the choice of the level at which to carry out the traffic captures has a bearing on the network coverage of a NIDS and, consequently, on the volume of traffic to be analyzed and on the set of cyberattacks that can be detected.

Data captured from root network nodes contains traffic coming from or going to a large number of network elements. Traffic captures at this level could allow Internet-related attacks to be detected by targeting as many network-connected machines as possible. However, some attacks would remain undetected because sometimes malicious activity goes horizontally, and does not involve the root nodes. NIDSs using root level captures need to process huge data volumes and must support extremely high throughput to avoid discarding traffic and to minimize detection delays. Also, due to the number of network elements they cover, they are easily affected by noise and perturbations, for example the connection of a new device on any segment of the network.

In contrast, the analysis of incoming and outgoing traffic corresponding to a small portion of the network, and captured from a lower level device, for example a datacenter switch, makes detectors more immune to noise and perturbations to occur at other network segments. Consequently, NIDSs using this data require more modest throughput levels. However, by design, their coverage is restricted to events occurring on a specific network segment.

Other scenarios are possible. Distributed captures aim to provide a trade-off between network coverage, performance and stability. Partial captures may be carried out at several network points and gathered at a central location, overcoming the drawbacks of using any of the two previously described approaches individually.

Data captured by network probes to feed NIDSs may have very different characteristics. Sometimes it consists of full copies of all data packets, such as those captured by a network sniffer. Another option is feeding the NIDS with traffic summaries, normally in terms of per-flow data samples, at the cost of having a limited view of the traffic that is traversing the network. We elaborate on these concepts in the following sub-sections.

### 3.1 Capturing Raw Packet Data

Communication networks move data units from one node to another. These units include information of a variety of network protocols, organized in levels or layers. Depending on the layer on which we focus, data units are called *frames* (LAN, link layer), *datagrams* (IP layer), *segments* (TCP/UDP transport layer) or *messages* (application layer). Often, *packet* is the generic name used for network data units. Unless otherwise stated, throughout this document we assume that we are dealing with IP datagrams.

A packet consists of a set of headers (control information), and the payload (actual data being transferred). Headers contain several control fields, required by network devices to handle the packet in order to be delivered to its destination. The payload is the data provided by the layers above –which includes header fields of those layers. Therefore, a single network packet includes several headers, each one belonging to a different layer of the TCP/IP protocol stack. Using a web example, the application header contains HTTP-related information, the transport header contains TCP-related information (for example, port numbers and different connection-related flags), the network header contains IP-related information (including IP source/destination addresses) and the data link header contains Ethernet-related information (including MAC source/destination addresses). Thus, a packet contains a large collection of fields of different natures: text, numbers, categorical information, etc.

All the mentioned protocol layers are necessary to move a message through the network. But not all the applications use the same protocol sets for this purpose. This means that not all the packets have the same fields, and even equivalent fields for two different protocols may be located at different positions within the packet. A long application message may be split into several sub-messages prior to transmission, due to limits imposed by link layer protocols. For instance, the Maximum Transfer Unit (MTU) of the Ethernet link-layer protocol is 1500 bytes. Such packets are reassembled later, at the destination. Thus, sending a single message may result in several packets. Also, some packets contain only control information, necessary for protocol operation, but without payload. Figure 2 shows the structure of a typical TCP/IP packet in an Ethernet-based network.

Packet capturing tools (a.k.a. network sniffers) such as *tcpdump* [21] or *Gulp* [22] can be used to retrieve and store raw network packets. They can run in a single computer, essentially capturing the packets arriving to/departing from that machine. A network device can be configured to mirror all packets to a single port, to which a sniffer is connected. This configuration allows traffic belonging to many machines to be captured but it has a bearing on the performance of the device, which is affected by the increasing number of devices to be monitored and the growing speed of modern networks. Thus, capturing raw data is a challenging task [23].

### 3.2 Capturing Flow-level Data

A *flow* is a collection of related packets that share a common set of attributes. These include network, transport or application header fields (IP addresses, IP protocol, TCP/UDP source ports, Type of Service...) as well as information related to the handling of packets on the device, such as physical input and output ports. Flows can be either unidirectional or bidirectional; in the later case they are also called *sessions*.

Flow-related features can be derived from raw packet data, as explained in Sections 4.1.2 and 4.1.3. However, the collection and reporting of flow-level data records is an increasingly common capability of network devices such as routers or switches. This feature aims to facilitate network management and supervision. Device manufacturers implement their own mechanism to gather and transfer this data, commonly adhering to the IPFIX standard [24], which was derived from Cisco's NetFlow v9 [25]. IPFIX-capable devices compute flow measurements in near real-time, without storing raw traffic, making use of sampling and accounting procedures. Sampling methods are used to take partial measures and approximate them to actual values when the number of connections is too high. Sampling parameters determine the behavior of the estimation process and play an important role in the extraction of useful characteristics for intrusion detection [26].

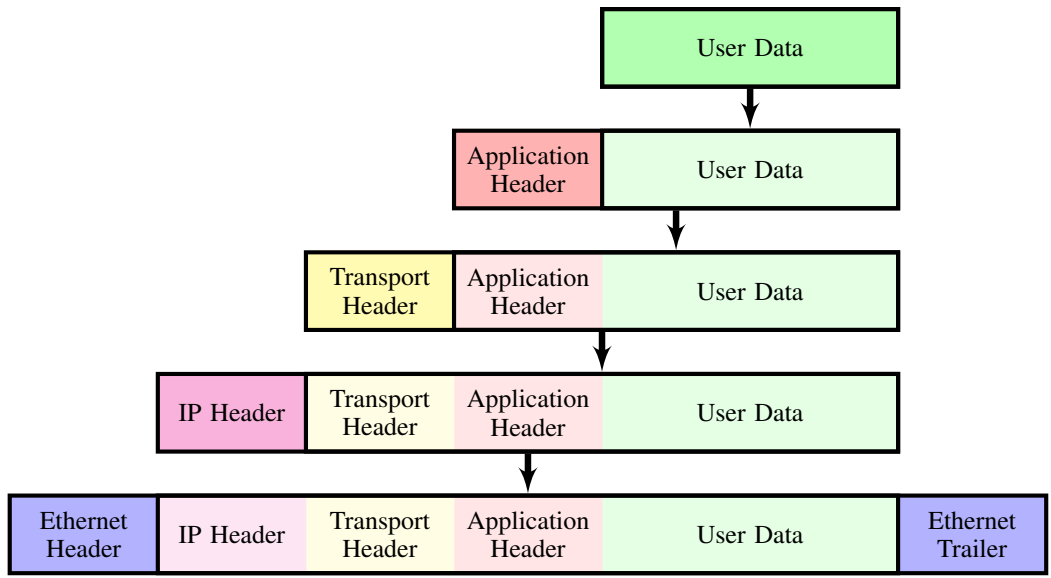

Figure 2: TCP/IP packet encapsulation: the information of upper level protocols is inserted into the payload at a lower level. Note that the application header is different for each application protocol, and that the transport header is different for TCP/UDP/ICMP.

Flow records computed in the device are exported to a *collector*, which stores them. Each record contains the basic information about a flow, plus additional variables, including: time stamps for flow start/finish times, number of bytes/packets observed in the flow, the set of TCP flags observed over the life of the flow, etc. Note that the payload, the actual end-user application data interchanged by means of the packets, is an optional field of flow records and, when present, it normally includes just a few octets. Therefore, flow captures may be far more compact than raw packet captures.

# 4 Preprocessing Stage: Derivation of Features

The preprocessing stage involves the derivation and transformation of feature vectors from the captured traffic data. As a result, a set of structured records, where each of them describe a different traffic observation (packet, connection, session) by means of a set of predictive or explanatory variables, is obtained. This phase also involves the process of labeling the records, i.e., assigning a response variable to each record which indicates its nature (attack or harmless).

## 4.1 Derivation of Explanatory Variables

The purpose of this phase is the extraction of structured records comprised of a set of *explanatory variables* or *features*. This is an important task of KDD because the detection capability of data mining methods highly depends on the information provided by input variables. In this section, we discuss different types of features, the procedures required to obtain them from network captured data and their usefulness for detecting cyberattacks.

Feature extraction can be done at different levels. We use the scheme proposed by Davis et al. [9] to describe those levels (using *flow* however, instead of *connection*-based aggregations) and the different classes of variables that are obtained at each level (see Figure 3).

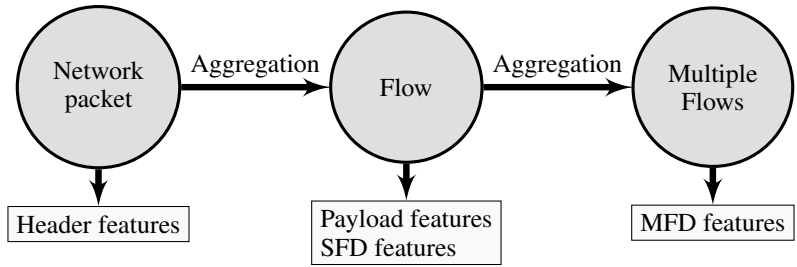

Figure 3: Feature derivation process

### 4.1.1 Packet Header Features

A raw packet can be converted into a feature vector by parsing the distinct set of header fields of the different protocols (from different layers of the TCP/IP stack) in use. Packet header features are useful to detect protocol misbehavior and low level cyberattacks such as spoofing (impersonation) attacks. However, many cyberattacks are only visible through the analysis of application-level data (headers and payload) that may be distributed along the payload part of several network packets, remaining inaccessible in a per-packet analysis.

Given the diversity in the structure of raw packets, and in the data types of the different protocol headers, the derivation of a common vector of features from this data [27] can be done in two different ways: (1) assigning missing values to those features that do not appear in a specific packet, or (2) removing fields that are not common to all protocols at the same layer. While the former leads to a more sparse representation of the data, the latter technique results in the loss of information that may be relevant for the detection of some attacks. In order to deal with both shortcomings, some authors proposed using different detectors, each one working with a particular protocol set [28, 29].

The main issue when working with packets (raw or converted into feature vectors) is not the preprocessing (parsing) procedure, which is straightforward, but the enormous throughput of packets sent to/received from a broadband link. Capturing, preprocessing and storing the packets in real time is a challenging procedure of growing difficulty [8, 23], which can be partially addressed through the use of different processing points working in parallel. Furthermore, the utilization of protocols involving encryption at the IP layer, such as IPsec [30], renders the obtention of higher-level headers and payload a challenging procedure (only IP and data link headers can be easily extracted). Given these difficulties, the use of higher-level data (see the following sections) constitutes a more practical alternative for real-world deployments [8].

### 4.1.2 Payload Features

Payload features are those obtained from the content of network messages, i.e., the data exchanged between applications. Payload features are useful to detect some cyberattacks such as injection or privilege escalation, which do not leave a specific footprint on packet headers. In such scenarios, where the attack vector is inside application-level messages, payload features have demonstrated their potential [31, 32].

As previously mentioned, for connection-oriented applications, a message can be split into a collection of packets (the number depends on factors such as the message size or the MTU of the data-link) before transmission. Those packets are reassembled at the destination, reconstructing the original message. Therefore, a preprocessing method which aims to extract meaningful features from application payload needs to implement this reassembly procedure.

Processing application messages is also challenging in terms of computing capacity, memory and speed. For example, a message may be really large (a connection may be used to send a multi-Gigabyte file), requiring enough storage during its transmission. Network size and speed may also stress the feature extraction mechanism, because storage and processing needs increase with the number of connections.

Encryption algorithms, used to provide security (confidentiality, integrity, authentication), make the extraction of payload features more difficult [33]. Efficient decryption mechanisms are essential in those scenarios in order to have access to plaintext (non-encrypted) payload. Such functionality typically requires additional processing infrastructure, such as TLS interception proxies that cut off encrypted traffic to decrypt and inspect it before being forwarded to the actual server [34, 35]. Drawbacks of these mechanisms lie in the increase of network latency and in the reduction of service security [36]. In practice, access of a NIDS to the contents of encrypted messages is unrealistic.

The extraction of payload features require additional preprocessing efforts to derive a common vector of features. These procedures are commonly based on those used for natural language processing [37]. They include vectors with the frequencies of characters [28] or bigrams (all subsequences of two consecutive symbols) [29, 31, 32, 38]. Note that resulting feature vectors are long, and may require additional dimensionality reduction procedures (performed in posterior phases of the KDD process).

### 4.1.3 Single Flow Derived Features (SFD)

SFD features are those that correspond to a collection of packets sharing any property, i.e., a flow. Packets are aggregated until a given event, such as the end of a TCP connection, is detected, or when a timeout triggers. Extracted SFD features summarize the properties of this aggregation of packets, by means of a set of statistical measures (such as byte and packet counts, as well as identification values). Common header fields also constitute features of interest. However, payload features are not considered part of SFD records. As a result of this derivation, a mixture of nominal and numerical variables for each record (flow) is obtained [39]. Tools to compute SFD features from captures of raw packets include Argus [40] and Bro [39]. However, a more common approach to obtain SFD features uses information computed directly by IPFIX-capable network devices.

SFD data sets are much smaller than raw data sets because payload is removed and common header features are summarized. This compactness is fundamental for many NIDSs, since it enables the operation in near real-time in high-speed networks. The consequence is a restricted ability to detect cyberattacks, limited to those that leave a footprint in the summarized statistics of a flow record (point anomalies) or occurring in an unexpected context (contextual anomalies).

Actually, lack of payload features in a context in which most information travels encrypted is not an issue. Regarding the feasibility to detect attacks which are visible through the analysis of the payload, some NIDS authors claim that the analysis of SFD features is a valid instrument [41]. However, their ability to tackle this problem is questionable due to the large number of false alarms reported in some papers [41–43].

A few NIDS papers making use of SFD features include descriptions about how they are derived [44–51]. They use a mixture of SFD statistics extracted from raw packet captures. Packets are aggregated using different header fields, such as the source IP address [17, 45, 52], or the source and destination IP address pair [17, 45, 52]. Computed features include byte and packet counts, average packet size, percentage of packets with a particular flag activated and average inter-arrival times, among others. Some authors extract and use more sophisticated features [50], such as the coefficients obtained from polynomial and FFT decompositions of measures such as inbound/outbound packet sizes.

### 4.1.4 Multiple Flow Derived Features (MFD)

MFD features are derived aggregating information pertaining to multiple flow records. Thus, they contain higher level statistics providing a more abstract insight about the traffic traversing the network. To compute MFD data, flows are commonly grouped using criteria such as "flows inside a time window $T$" or "last $n$ flows". Then, sets of statistics are computed for each group [53]. Note that packets within a flow share many properties (SFD features), while flows aggregated in a group may not share any common characteristic beyond being in temporal and, sometimes, spacial proximity.

MFD features are useful to identify attacks that spread their activity over several connections, such as (D)DoS attacks or probes [45]. However, when used as the single source of data for the detector, it is difficult to trace the source of the attacks, due to the absence of packet or flow identification fields [53]. Furthermore, computing these features has an extra cost in terms of processing and memory resources.

NIDS papers proposing the use of MFD features aggregate flows using time windows or flows comming from/going to the same host as aggregation criteria. From those aggregations, counts of bytes, packets and flows are obtained in [45]; entropy values are calculated in [6, 53, 54]; and covariance matrices are computed in [55, 56]. Other authors [57], used instead a fixed-width window of flows as their aggregation criteria to subsequently apply the Discrete Wavelet Transform [58] and compute new MFD features.

### 4.2 Class Labeling: Derivation of Response Variables

Response variables, or labels, are defined to explain the actual nature of a data record: non-malicious or malicious. In a fine-grained scenario, labels may even represent the specific attack type. These variables are usually required for evaluation purposes, to compare the output of a detector with the actual label. Also, the availability of labeled records is mandatory for training some supervised data mining approaches such as classifiers (see Sections 7.1 and 8).

Class labeling can be accomplished through different mechanisms, used alone or in combination. Manual labeling is done by analyzing each record in the dataset, together with its context, and deciding whether a record is malicious or not. This task must be carried out by experts because the identification of malicious patterns requires specific domain knowledge. Thus, it is a very time consuming procedure but results in very reliable labels.

To automate the labeling process, external tools can also be used. Normally, the output of several tools, specialized in matching signatures for known attacks (misuse-based NIDS), is used in conjunction with databases of cyberattacks and lists of blacklisted hosts [59, 60]. However, this kind of labeling is prone to errors, with either harmless records flagged as malicious, or attacks that go undetected (false negatives). In such cases, it is commonplace to tag as *background*(instead of *harmless*) the traffic that is not identified as malicious [59].

Yet another way of generating labeled datasets is through a synthetic environment in which well identified cyberattacks are injected into a network at specific, and well controlled, times. The injection schedule is used to do the labeling, by matching attack times with the time stamps of network records. Scheduled labeling ensures the identification of all malicious records while keeping the labeling cost under control. A disadvantage of these process is the lack of realism in the scenario, as both the normal and the malicious traffic is artificially generated.

Note that labels may be (1) unbalanced, because in a network there is more harmless traffic than attacks, and not all attacks are equally common; and (2) unrealistic when, in order to make it more balanced, the proportion of attacks is artificially high [61]. Both alternatives have a bearing on the performance of data mining algorithms [62]. Several authors modify the datasets in order to have a "fair" label representation [61, 63–69].

### 4.3 Publicly Available Datasets for NIDS Research

As explained in the previous sections, capturing and preprocessing data are tedious, non-trivial processes, particularly if labeled records are required. For this reason, most researchers use publicly available data sets specifically designed to help building, testing and evaluating NIDSs. They are ready to use, and have an additional advantage: the possibility of comparing the performance of different NIDS proposals with a common yardstick.

Table 3 shows a summary of popular NIDS datasets, the formats and features available in each one, as well as the labeling method used by their creators. It can be seen that most datasets provide already preprocessed records, and only a few of them provide the raw captures from which higher-level (SFD, MFD) features were derived. This is a clear drawback and, ideally, all datasets should include the raw captures,allowing researchers to obtain those header, payload, SFD and MFD features that they find more adequate –instead of being limited by those already included in preprocessed data. Some datasets obtained in synthetic or simulated environments use scheduled or manual labeling methods. In datasets obtained from real networks, labels are normally assigned using external tools.

A deeper investigation about the generation process shows that some datasets [59, 60, 70–74] are captured in real networks during the operation of real users. Others instead [75–77] contain traffic captured in synthetic environments, designed with the aim of being as realistic as possible, but with limited infrastructure; hence, these datasets are characterized by including a small number of hosts (sources/destinations of traffic) and limited traffic heterogeneity, which may be easy to model – resulting in biased and scenario-dependent detectors. Moreover, traffic generation methods include programs that try to mimic the operation of the users [75–77], but are not real users. In general, the adequacy of synthetic datasets is in question as they do not truly reflect the operation of actual networks in terms of size and variability of user behavior (legitimate or malicious) [59].

| Dataset | Raw captures | Payload features | SFD | MFD | Labeling method[1] | Duration | Year of Captures | Continuous capture | Real scenario | #articles |
|---|---|---|---|---|---|---|---|---|---|---|
| DARPA 1998 [70] | ✓ | | | | MS | 9 weeks | 1998 | YES | YES | 6 |
| DARPA 1999 [71] | ✓ | | | | MS | 5 weeks | 1999 | YES | YES | 5 |
| KDD-CUP 99[2] [72] | | ✓ | ✓ | ✓ | MS | 9 weeks [1] | 1998 | YES | YES | 36 |
| NSL-KDD[2] [73] | | ✓ | ✓ | ✓ | MS | 9 weeks [1] | 1998 | YES | YES | 13 |
| Kyoto2006+ [74] | | | ✓ | | E | 34 months | 2006-2009 | YES | YES | 3 |
| MAWILab [60] | ✓ | | | | E | 15min daily traces | 2001- | NO | YES | 3 |
| ISCX-2012 [75] | ✓ | ✓ | | | S | 1 week | 2010 | YES | NO | 3 |
| TUIDS [76] | ✓ | | ✓ | ✓ | S | 1 week | 2011 | YES | NO | 3 |
| UNSW-NB15 [77] | ✓ | ✓ | ✓ | ✓ | E | 31 hours | 2015 | NO | NO | 2 |
| UGR16 [59] | | | ✓ | | ES | 19 weeks (132days) | 2016 | NO | YES | - |

[1]M=Manual labeling, E=External resources, S=Scheduled attacks
[2]Derivatives of DARPA datasets

Table 3: A summary of datasets available for the evaluation of intrusion detection systems.

Another issue of these datasets is the short duration of captures, commonly lasting only a few weeks. Networks are changing environments where traffic patterns, harmless and anomalous, change throughout time and rarely manifest long stationary periods. Short duration captures over intermittent periods limit the correct evaluation of the adaptability of the models when the network changes. Some commonly used datasets were created more than a decade ago and, therefore, contain outdated traffic, meaning that old protocols and services are present, while current ones are not. Also, ciphered traffic is omitted. In summary, the included traffic patterns do not represent current networks. At any rate, being recent does not guarantee being representative, especially if the dataset is synthetic.

The most used datasets in the literature (see Table 3) are the DARPA datasets and their derivatives (KDD-CUP 99 and NSL-KDD), which have been largely criticized due to statistical issues [73, 78–80]. Despite the fact that better alternatives have become available in the last few years, DARPA datasets and their derivates are still used to develop and test intrusion detectors [81].

## 5 Preprocessing Stage: Data Cleaning and Feature Transformation

The main characteristic of the feature vectors (records, samples) obtained from the previous stage is their heterogeneity. They include discrete (categorical and numerical) and continuous (numerical) features. As some machine learning methods cannot deal with this heterogeneity, it is common to transform the features to obtain homogeneous records (all numerical or all categorical). Also, data mining algorithms are very sensitive to noise in the data caused either by outliers or errors during the data capturing process. The procedures used during this second preprocessing stage involve the derivation of new features, the modification of existing ones and the elimination of noisy samples, all in order to build a "clean" dataset.

Unfortunately, little attention has been paid to this stage by NIDS authors. Most papers do not provide enough details about the exact cleaning and transformation procedures they perform, making their analysis difficult and hindering the reproducibility.

### 5.1 Noise Reduction

Data capture tools may be susceptible to errors during traffic processing. Such errors, commonly denominated noise, are reflected in the data in different forms. Noisy data affects the learning capacity of data mining detectors negatively, making them prone to failures [82]. Thus, noise reduction procedures aim to increase the effectiveness and accuracy of data mining algorithms. Noise can be reduced by deleting those samples showing higher distortions with respect to the majority of elements in the dataset, either in a single feature with extreme or missing values, or in several features. Note that, as we describe in the remainder of this section, other transformation methods also contribute to the reduction of noise in an indirect manner.

Only a few NIDS papers explicitly describe the noise reduction techniques they use. Examples are [83] and [52], where outlier removal is done before feeding the dataset to its respective detectors.

The former method uses the Mahalanobis distance to rank samples and remove the 0.5% with the largest values, whereas, the latter technique uses an interval for every feature. To do so, limits away from the mean value are calculated taking 10% of the distance between the maximum and minimum feature values.

## 5.2 Transformation from Categorical to Numerical

There are different approaches to convert a symbolic feature into a numerical representation. A naive conversion is called *number encoding*. It keeps the dimensionality of the dataset, assigning a number to each different symbol. The method is useful to simplify value comparison and to reduce memory utilization. However, while distance or order relationships are inherent to numerical data, this is not necessarily true for categories encoded numerically. Therefore, this solution may lead to erroneous conclusions when using certain data mining algorithms. This method has been used for NIDS in [84],[85] and [86].

To avoid the shortcomings of number encoding, *one-hot* encoding adds a new binary feature (dummy variable) for each categorical value in the dataset, and assigns a positive value (1) to those observations sharing the category represented by the binary feature, and a zero value (0) otherwise. This approach has been commonly used in NIDS proposals to enable the use of distance calculations [64, 87–89] without the need to use a heterogeneous measure. Other NIDSs, whose models cannot work with numerical values, also use this transformation [57, 90–96].

A variant of the previous method is called *frequency encoding*. Instead of assigning a value of 1, each feature value receives a weight according to the frequency of appearance of that value in the data. This method was firstly proposed for NIDS in [65], because it is useful to work with distances. Based on this first approach, frequency encoding was also used later in [68].

## 5.3 Feature Scaling

Feature scaling, also called feature normalization, is used to limit the range of values of a numerical variable. It has been demonstrated that scaling procedures improve the convergence time of optimization algorithms, as they reduce the solution search space without negatively affecting the accuracy [97]. Applying some scaling methods may also reduce the skewness in the data [98], as well as the bias caused to some data mining algorithms when using variables with large values [99, 100], such as those based on distance measures [83].

The main drawback of scaling methods is that they may require knowledge about the value ranges of the features. This is problematic in dynamic scenarios such as streaming contexts, where the range of the variables is not known. Additionally, for features where most values are within a very narrow value range, feature scaling methods result in a reduction of the differences between (transformed) values when outliers are present, disturbing the operation of some data mining algorithms [93]. Noise removal and discretization (which will be explained in the following sections) methods may help reducing this side-effect.

The use of feature scaling has been documented in several NIDS articles. In [53, 101], mean normalization is carried out to scale numerical values before the application of principal component analysis (PCA, see Section 6.1.3). The method uses the mean and the difference between maximum and minimum values to center the data. A logarithmic transformation is used in [85, 92] to reduce the variability of features with wide value ranges. Min-max normalization uses the minimum and maximum values of a variable to adjust its range to values between 0 and 1; this method is applied in [86, 94, 102] and [90–92] to reduce the training time of the models used for detection. The same approach is used in [52, 66, 93] to avoid the bias caused by features with wide value ranges in distance-based algorithms. Standardization takes the sample mean and the standard deviation to transform a feature to have zero mean and unit variance; it is used in [64, 65, 88, 89, 103] to improve, again, the behavior of distance-based algorithms. Both, standardization and min-max normalization methods, are used in [104] to improve the performance of the optimization algorithm which estimates the parameters of their detection model.

## 5.4 Feature Discretization

Discretization is used to transform a continuous feature into a discrete variable [105]. Although the use of these techniques is largely discussed in the machine learning literature [106], it is not that ubiquitous in the NIDS literature. Discretization works by selecting split or cut points over the continuous space of a variable to create subdivisions (intervals) of this space. Then, each continuous value is replaced by the corresponding interval identifier.

Variables are discretized with the purpose of increasing the learning speed of some detection algorithms, such as those based on induction rules [107]. In decision trees, the use of discretization improves the effectiveness of the learning procedure, resulting in smaller and more accurate trees [108]. Discretization also reduces the complexity of algorithms based on the Bayes theory (by using summation instead of integration terms), while increasing the accuracy of resulting classifiers [105]. Furthermore, it is useful to remove noise from data and to reduce the repercussion of outliers [93], at the cost of some information loss.

The simplest discretization algorithms are Equal-Width (EWD) and Equal-Frequency discretization (EFD). EWD splits the original feature range into $k$ intervals of equal width; EFD instead creates as many intervals as needed, in such a way that each interval contains the same number of $k$ samples. A similar method uses K-means clustering, which splits the feature range in $k$ intervals and replaces the original value of a feature with the centroid of its corresponding interval. In all cases, the number $k$ of intervals (or elements per interval in EFD) must be selected by the user. For EWD and K-means small values of $k$ may remove information useful to explain the data, whereas large values of $k$ slow down algorithms and may be inadequate to properly remove noise in the data. The opposite is true for EFD.

Other popular, more complex algorithms used in NIDS are Entropy Minimization Discretization (EMD) and Proportional k-interval discretization (PKID), which avoid the need to set the number of intervals. EMD recursively selects as interval cut points those elements with minimum values of information entropy with respect to the class [109]. PKID [110] selects a number $k$ of intervals in such a way that each interval contains a number of samples that is equal to $k$ (the number of bins), trying to find a trade-off between the choice of $k$ and the sparseness of the data.

To the best of our knowledge, the use of discretization techniques in the NIDS literature has not been largely documented, but there are some exceptions. The authors of [111] show how the use of different discretization methods (EWD, EFD, PKID and EMD) has an impact on the relevance of the features chosen by feature selection methods (see next section), and therefore on the detection performance of their NIDS. Best results were obtained using EMD and PKID. Influenced by this previous work, in [112] these methods are used to transform continuous variables into discrete values before training a Bayesian detector. EMD is also used in [47] to speed up the building of a similar detector.

Some NIDS approaches based on Decision Trees [47, 55, 66, 93, 113] use discretization indirectly as it is embedded into the learning algorithm. These classifiers use a discretization method based on information gain and the minimum description length (MDL) principle, splitting a continuous variable space by selecting the cut point which obtains the maximum information gain from the split [108].

To conclude this section, a special case must be considered when processing data on-the-fly and the ranges of the features are not previously known. Online discretization methods make approximations of feature value ranges and compensate them as new data arrives [114, 115]. A solution at training time to this problem is proposed in [93], where the authors cluster each dynamic-range feature into $k$ groups applying K-means. After that step, $k$ Gaussian distributions are obtained (one for each group). Next, during the operation, every dynamic range feature is substituted by $k$ new features obtained using the Gaussian distributions and containing the probability of the original feature value pertaining to each group.

## 6 Data Reduction

In the previous sections we have discussed the derivation of an exhaustive set of features to form the dataset, which later will be used to build the detector. This dataset may be huge in terms of the number of samples, and also in terms of the number of features per sample. In this context, it has been shown that using the full feature set extracted from traffic is ineffective for intrusion detection, due to the inclusion of irrelevant variables that contain redundancies and hinder the learning capacity of data mining algorithms [116].

In this section we discuss methods to reduce the dimensionality of the dataset with the aim of: (1) reducing the computational cost and (2) improving the accuracy of intrusion detectors. We describe some techniques that reduce the number of variables in the feature set, as well as others that reduce the number of samples (records) in the dataset.

### 6.1 Feature Dimensionality Reduction

Feature dimensionality reduction procedures try to cut down the number of features considered by the NIDS by discarding useless information or reducing redundancy. There are two main alternatives to do this: removing explanatory variables, or projecting the original explanatory variables onto a lower dimensional space.

#### 6.1.1 Manual Feature Removal

The simplest feature reduction method is the manual removal of "irrelevant" features. It requires a profound expert knowledge of the domain, as those features are excluded for the next KDD phases [117].

As previously mentioned, many data mining algorithms cannot work with some types of attributes (categorical, numerical). To deal with this issue, many authors simply opt to remove features instead of applying preprocessing methods (see Section 5). For example, in [55, 83, 103, 118–121] all categorical variables are removed because their data mining detectors are unable to deal with non-numerical data. Note that this removal may result in important information loss [118].

Manual exclusion of features based on other criteria has also been described in some NIDS papers. In [55, 56] only MFD statistics of the dataset are used for DoS detection, discarding those that are not desired. In other proposals, constant features are removed as they do not provide useful information [49, 68, 90]. In [47] features such as IP addresses of attacking machines and attacking times are removed, as they reflect information clearly related with the simulated environment where the dataset was obtained from.

#### 6.1.2 Feature Subset Selection

Feature Subset Selection (FSS) techniques select a subset of the whole feature set that is assumed to be the most relevant to solve the problem at hand. FSS methods reduce the cost of data mining models, help to prevent model over-fitting [122], and enable a better understanding of the data because redundancies are removed [123]. FSS methods can be classified into three main categories: wrapper, filter and embedded methods.

*Wrapper FSS* approaches measure the benefit that the inclusion of a given feature (or set of features) has over the performance of a predictive model. This can be seen as an optimization procedure that looks for the feature subset that maximizes the predictive capacity of the model. Different methods to select feature subsets have been described for NIDS, including sequential methods [38] and those based on meta-heuristics [47, 124, 125]. The main drawback of wrapper FSS methods lies in their computational cost, much higher than that of other FSS methods as it entails several training-testing model iterations to fully evaluate the feature set.

In contrast to wrapper methods, results of *filter FSS* methods do not depend on the predictive model used in the next KDD phase. The relevance of features is computed through an evaluation of their characteristics using correlation, mutual information, consistency, variance or similar metrics [126]. Normally, filter FSS methods are computationally cheaper than wrapper methods, and more appropriate for datasets containing unlabeled data or a large number of features.

Correlation-based Feature Selection (CFS) methods [127] choose those features that are highly correlated with the response variable (the label) while showing a low correlation with the remaining features. To compute such correlation, Pearson's correlation coefficient is commonly used, although this coefficient is only able to express linear relations. Other correlation-based filter methods use

Mutual Information (MI) (based on entropy), which is able to measure complex relations [128], but can be applied to discrete features only. CFS methods based on Mutual Information have been used for NIDS in [85, 86, 102].

Consistency-based methods evaluate the constancy of feature values with respect to class labels [129]. As such, features with stable values in samples belonging to the same class are selected. For NIDS this method has been used in [111, 112] in conjunction with two CFS methods, one based on Pearson's correlation coefficient and the other one based on MI [130].

Analysis of Variance (ANOVA) tests [131] decompose the variance of features among linear components, which can be used to compute the variance of a feature with respect to others. Then, redundant features containing similar characteristics to others can be discarded. The NIDS proposal [57] uses ANOVA to select those features that provide non-redundant information.

A combination of filter and wrapper approaches is proposed in [86] for NIDS. The method uses a MI-based ranking, such as those used in filter methods, to determine the order in which features are subsequently entered into a wrapper FSS method.

Finally, *embedded FSS* approaches are either filter or wrapper methods integrated into prediction algorithms, and applied as part of the training phase of those algorithms. They are dependent on the particular data mining algorithm used, usually trees, as each training step comprises both ranking relevant explanatory variables and generating a model using the top-ranked ones [122]. For example, C4.5 [132] and C5.0 [108, 133] decision tree builders are used in [47, 66, 94, 134] and in [55], respectively, to generate intrusion detectors. Both algorithms have a built-in filter FSS method that weights features using their Gain Ratio (an improvement over MI used to avoid bias towards variables with higher cardinality). Other authors use Random Forests for intrusion detection [48, 63, 135], which includes a wrapper FSS as part of the learning process. The feature relevance metric is based on the misclassification rate produced by the permutation of the values of the input variables [136].

### 6.1.3 Feature Projection

Feature projection techniques do not select a subset of the original feature set. Instead, they are used to obtain a projection of the features into a lower dimensionality space. It is a form of compression, based on the use of statistical and mathematical functions that find linear and non-linear variable combinations that retain most of the information of the original feature vectors. Because projections do not contain the original values extracted from traffic samples, their main shortcoming lies in the lack of explainability of detected attacks.

Principal Component Analysis (PCA) [137] is one of the most common dimensionality reduction procedures. PCA applies a linear transformation to the original explanatory variables to map them into a space of linearly uncorrelated components. Dimensionality reduction is performed by selecting a number of components smaller than the original number of features [138]. Ideally, data provided to PCA must be scaled and free of outliers to avoid giving more weight to features with wider value ranges [83]. The main shortcoming of PCA is its inability to capture complex, non-linear relations between features. Several NIDS works use PCA to reduce the feature dimensionality of their data before training their models. In [139] several training and testing executions are performed using PCA in order to select the number of components that result in the highest performance. The authors of [88] select those components that best discriminate between normality and attack, as indicated by Fisher's Discriminant Ratio.

In contrast to PCA, where the extracted components are orthogonal and ordered by the variance they retain, Independent Component Analysis (ICA) separates a multivariate signal into a set of statistically independent non-Gaussian components or factors of equal importance [140]. In NIDS, this method is used in [46] to obtain a representation of the data in which only those independent components that better characterize the malicious class are kept.

Auto-encoders, a type of Neural Networks (NN), have also been proposed for feature projection. They try to reproduce the input values (the original features) at the output layer. Dimensionality reduction is achieved by projecting the input onto a smaller space in the intermediate layers. Auto-encoders can model complex relations in the input data, not limited to linear relations [141]. Different auto-encoder methodologies have been proposed in the NIDS literature, including symmetric [90, 142] and non-symmetric [143] NN architectures, sparse auto-encoders[1] [51], and auto-encoders based on self-taught learning techniques [144]. Four different auto-encoder models are used in [96] to determine the best NN-based projection method for NIDS.

Note that dimensionality reduction techniques can be used instead of FSS, but also after doing FSS, thus getting a very compact representation of the data. This approach is followed in [139], where PCA is applied to obtain a reduced projection from the top ranked features returned by a filter FSS method based on information gain.

### 6.2 Sample Dimensionality Reduction

The cost of training most data mining algorithms depends heavily on the number of samples used to train the models. Thus, techniques that reduce sample dimensionality are sometimes applied with the aim of accelerating the learning step of those algorithms, either removing duplicate or too similar samples, or selecting representative samples that are used instead of the whole dataset[2].

In the context of NIDS, the authors of [87] use a tree-based clustering method to reduce the number of samples used to train their model. Instead of training with all the instances, the cluster centroids are used as representatives, thus achieving a significant reduction in the sample dimensionality. A similar method is used in [118, 145] to reduce the cost of performing the distance calculations required by their ML detectors.

## 7 Data Mining Stage

Data mining is the core stage of the KDD process, where the preprocessed and reduced data is fed into different algorithms whose objective is to identify patterns in the data and, ultimately, detect attacks. Due to its core role, most NIDS surveys (such as [10, 12]) focus almost exclusively on this stage. In fact, the authors of those surveys categorize NIDS proposals using the type of data mining algorithm used as the main criterion. In this survey we discuss this stage using a slightly different viewpoint, and present a taxonomy of NIDS methods based on two criteria: the detection goal and the learning approach.

The first criterion categorizes NIDS methods into three types (misuse-based, anomaly-based and hybrid systems) based on the meaning they assign to the concept of attack and the goal of the detector. The second criterion classifies NIDS proposals according to how they train or build the detector: in batch mode or incrementally.

### 7.1 Misuse-based, Anomaly-based and Hybrid Detectors

This nomenclature (misuse, anomaly) has already been used in previous surveys such as [12, 14] and, even if the definitions of these terms given in the literature are not always identical, the different versions do have some common ground. In general, most authors have defined misuse-based systems as NIDSs designed to detect well-known attacks using either a knowledge-base of attacks [8], rules written by domain experts [9], attack-signatures [7, 12, 14, 146] or attack-patterns [17, 27]. The authors do not specify the exact meaning of *signature*, *rule*, *pattern* or *knowledge-base*, but clearly some definitions refer to very rigid systems (signature, rules, expert knowledge), while others describe misuse-based NIDS as more flexible systems (patterns). In most works, the authors do not state clearly how this kind of identifying information can be extracted from data.

In the case of anomaly-based detectors, the definitions are vaguer and, oftentimes, with slight differences. These NIDSs have been defined as systems whose objective is to detect *exceptional* patterns [10], patterns that *deviate from other observations* [11], patterns that *deviate from the normal* [7–10, 12, 14, 27] or patterns that *deviate from the expected* traffic [5], and also as systems *with knowledge of normal behavior*[146]. In this line, in many of the mentioned surveys, their authors state that an anomaly-based NIDS attempts to obtain profiles of the normal traffic, in order to compare the new incoming data with those profiles, identifying observations that deviate significantly from normality.

---
[1]These projection methods do not result in a reduced projection, instead they produce a sparse representation where the projected data is larger than the original feature set.

[2]These techniques are not to be confused with class balancing, whose aim is to reach a proportional, and maybe unrealistic, distribution of samples between the classes.

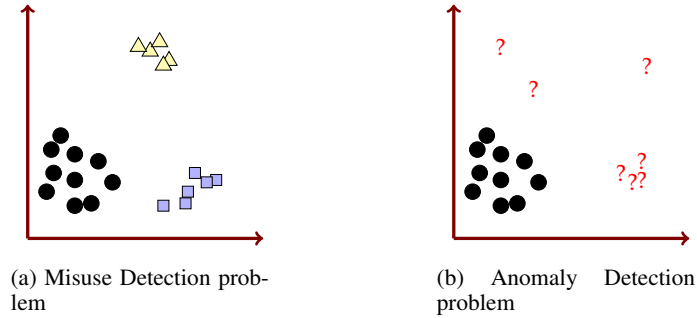

(a) Misuse Detection problem

(b) Anomaly Detection problem

Figure 4: Visual description of the misuse and anomaly detection problems. In misuse systems, both the normal data (black circles) and the attacks (yellow triangles and purple squares) have specific characteristics and, thus, form classes. In anomaly scenarios, normal observations (black circles) are assumed to follow characteristic patterns that constitute well-defined classes, while abnormal, maybe malicious, patterns (red "?" symbols) may or may not have enough entity to form a class.

|  | Do the attacks form classes? | What is modeled? | What is the response obtained from the NIDS? |
|---|---|---|---|
| **Misuse** | Yes | Attacks and Normality | Normal observation vs. attack (or type of attack) |
| **Anomaly** | Not necessarily | Mainly normality | Normal observation vs. Abnormal observation |

Table 4: Differences between misuse and anomaly detection from a data mining perspective.

Often, different surveys classify the same NIDS differently, as evidence of the absence of a clear-cut categorization. For example, proposals [32] and [63] are categorized as anomaly-based and hybrid, respectively, in [10]. However, in [12] both proposals are catalogued as misuse-based. In recent surveys, the inconsistency of these definitions has lead the miscategorization of NIDS using some data mining methods, such as classifiers [10, 14, 15], identifying them as anomaly-based approaches although they present limitations to detect new and unknown classes of traffic [17, 147].

To avoid further misclassifications, we start laying the foundations for these definitions, providing more specific and up-to-date descriptions of misuse and anomaly based approaches. We focus on three aspects that depart from ideas and concepts of the data mining area, particularly from the area of anomaly/outlier/rare event detection [148] (see Table 4). Any combination of the characteristics shown in this table will result in a hybrid detector.

- **Misuse-based detectors** understand attacks and normality as categories/classes within the data (see Figure 4a). They depart from the assumption that an attack is a minority category with specific patterns/characteristics that can be learned using the training data. Misuse-based systems require fully labeled data (all attacks must be correctly identified) at training time. Once a misuse-based system is built, the response obtained given a new observation is its correspondent or potential class according to the learned characteristics. Therefore, detection ability is limited to the set of attacks and/or normality classes modeled.
- **Anomaly-based detectors** assume that attacks can form categories with specific characteristics, but they can also be isolated observations (outliers) (see Figure 4b). With this in mind, these detectors regard that an attack is anything deviating from normal traffic and, therefore, they focus on modeling normality. Anomaly-based NIDS do not require attacks to be explicitly identified in the training data, and the output they produce during operation is vaguer than in misuse systems, since the returned information only states that an observation deviates sufficiently from normal traffic.

This categorization of NIDSs is closely related to the degree of supervision required by the data mining algorithm applied. Classical supervised classification algorithms depart from a training dataset in which the instances are labeled as attack (or type of attack) or normal traffic. Their goal is to learn a predictive model from this data, which is then used to identify attacks in new incoming traffic [12]. As such, they are typically associated with misuse-based systems. On the contrary, unsupervised learning algorithms depart from a dataset of instances that are not labeled at all. Their goal is to identify the natural groups that underlie in the data [149] and, thus, are typically more related to anomaly-based systems.

In anomaly-based NIDS we can also find different types of weak supervision [150], where some kind of uncertainty is included in the labeling. To the best of our knowledge, two types of weakly supervised learning approaches have been applied to NIDSs: semi-supervised learning and one-class classification. While in semi-supervised learning [151] the data is partially labeled, in one-class classification [152] only some of the instances of one of the classes present in the data are labeled. These algorithms do not require as many labeled instances (which are difficult and costly to obtain, see Section 4.2) as fully supervised approaches, and need shorter training times. Furthermore, generally they have better accuracy than totally unsupervised models.

In the following subsections we describe NIDS works from each of these categories, commenting on the advantages and disadvantages of each approach. We also discuss the degree of supervision required, or typically assumed, by the different detection methods.

### 7.1.1 Misuse-based NIDS

As mentioned above, misuse-based systems normally leverage classic supervised classification methods. These algorithms have evolved and improved substantially in the past few years and, consequently, they are able to yield very reliable predictions when fed with enough, good quality data. The main advantage of misuse-based systems is that they are able to accurately model the patterns and characteristics of well-known attacks present in the training set, obtaining low rates of false alarms (false positives). However, most of these techniques have important limitations when dealing with new (zero-day) or mutated attack patterns [47], because they are not designed to identify new classes. Thus, they require continuous updates to extract new attack patters, making them hard to maintain. Also, remember that this type of algorithms require a labeled training set, which is costly and difficult to obtain in real scenarios [153] (see Section 4.2).

The main difference between the various misuse-based systems proposed in the literature is the specific supervised classification algorithm used. Neural Networks (NN) are rather popular and, in NIDS, there are approaches ranging from the simplest network, the Multilayer Perceptron (MLP), [51, 57, 90, 154], to complex Extreme Learning Machines (ELM) [49] and Recurrent Neural Networks (RNN) [69, 91, 92].

Other less frequent approaches include Support Vector Machines (SVM), based on finding the maximum margin hyperplane that separates the classes, [38, 57, 86, 117, 125], k-Nearest Neighbor (kNN) approaches, based on distance calculations [65, 89, 155], the Naive Bayes algorithm, based on probabilistic theories [44, 47, 57, 111, 112, 125, 156] and Decision trees, based on constructing tree-like structures that output decision rules [47, 55, 66, 94, 111, 113, 125]. As a more rare approach, a supervised probabilistic variant of the Self-Organizing Maps (SOM) algorithm [157] (which is an unsupervised clustering algorithm), has been applied to NIDS in [158] and [88], assigning to each cluster the most frequent label in its allocated data during the training phase.

The combination of supervised classification methods to form ensemble models has also been proposed for NIDS, since it reduces bias and improves classification accuracy [136, 159] compared to using only one model. The most common ensemble method is the Random Forest [17, 48, 63, 94, 143], which is based on combining many decision trees. However, other heterogeneous ensembles have also been proposed: combinations of Decision Trees and SVMs [134]; ensembles of Neural Networks, SVM and Decision Trees [94]; or even combinations of SVM, kNN and MLP classifiers [139].

We already mentioned that misuse approaches are only able to recognize what is learnt during training. In this context, in order to provide some flexibility and to avoid overfitting to the training data, some supervised methods include regularization or generalization terms [160]. An alternative approach to provide more flexibility to these methods and attempt to generalize better is proposed in [118]. This misuse-based detector is a variant of kNN, a distance based classifier, that incorporates a genetic approach. This genetic component makes every new sample evolve to perform its classification using a clustered representation of the training data.

We have seen many distinct methods that are used for misuse detection. A comparative analysis of them is not easy due to the lack of a common framework. However, some general conclusions can be extracted through an analysis of the algorithms applied. Among the simplest approaches we can find DT and Bayesian methods, which involve a straightforward learning mechanism. In contrast, NNs and SVMs need a thorough parameter tuning. Most algorithms are costly to train and perform better when they are fed with large datasets. kNN is costly in the detection phase because, to obtain the label for a new sample, it requires the computation of the distance between that sample and all those of the full training set. SVMs are good choices for learning from wide datasets (those with few records and many features). Regarding model interpretability, DTs are the most explainable, while the opposite holds true for NNs.

### 7.1.2 Anomaly-based NIDS

These detectors assume that an attack is anything that deviates from normal traffic, and thus, focus mainly on modeling "normality". Deviation of new observations from this normality has been measured in many different ways. As previously mentioned, while misuse-based detectors are commonly based on supervised learning schemes, we can find supervised, weakly supervised and non-supervised approaches for anomaly detection.

A main advantage of anomaly detectors is that they should be able to identify unknown attacks (mutated or zero-day attacks). Labeled data is not a requirement for such systems because normal profiles can be modelled using unlabeled or partially labeled network datasets, which are expected to have a low proportion of attacks. A main drawback of these detectors is that, when deployed in real scenarios, they usually produce a high number of false positives, often related to noise or events that are not attacks (e.g., the connection of a new server in the network). Therefore, anomaly-based NIDS typically require companion methods to explain/categorize the alarms [161].

An anomaly-based NIDS identifies as attacks those observations that are very different from the normal, majority traffic. As such, a threshold must be defined to determine if the level of deviation is enough to trigger an alarm. Anomaly-based NIDS proposals differ in the data mining algorithm chosen, as well as in the way deviation is measured. Let us focus first on unsupervised approaches.

The first and most simple unsupervised anomaly-based systems are based on the extraction of frequency profiles from the patterns observed during training. These detectors assume that attacks leave as footprints patterns that do not appear as frequently as normal patterns. For example, in [28, 29, 64, 162, 163] the authors flag infrequent values in traffic features traffic as malicious. More sophisticated is the approach followed in [119]. Traffic of similar characteristics is first clustered, and the frequency of transitions between clusters is analyzed. Attacks are identified when uncommon transition patterns appear.

Other approaches are not based on frequencies, but on modeling the correlation structure of the normal data and assuming that attacks follow a different correlation structure. Two main approaches can be identified here: using correlation features directly, or using subspace representations of the correlation features. Examples of the first approach can be seen in [56] and [103]. In the first case the mean of the covariance matrices of normal data is computed. Later, records that deviate from this profile beyond a threshold are flagged as attacks. The second proposal computes the area of the triangle formed by the projection of any pair of variables in the Euclidean 2D space, which is used as an indicator of their correlation. The system generates a normality model by computing the Mahalanobis distance between all normal triangle area records, to obtain the parameters of a univariate Normal distribution. After that, a threshold for admissible normality consisting of $n$ times the standard deviation of the normality profile is set.

Subspace representations of the correlation features by means of projection methods (PCA) have also been used for anomaly detection. For example, in [83] the authors use PCA to extract the major and minor components of normal data. The $k$ major components are those capturing most of the variance.

Incoming records with normal correlation structure and extreme values are flagged as anomalous by means of the reconstruction error of their projection using the major components. Similarly, uncorrelated records are detected using the minor components. Other approaches that use PCA to exploit the correlation structure of the data are described in [53] and [6]. Both apply PCA to entropy features, considering the minor components as indicators of how well the normality is represented.

In [53], the residual projection of the data obtained by means of the minor components is used to identify anomalies with a high reconstruction error. In all these cases, the number $k$ of major components is selected at training time. In contrast, in [6], the authors use the angle between the PCA projections of two contiguous non-overlaping windows of data to obtain an anomaly score and, to do so, the number of major components $k$ is tuned at run time, selecting the number that minimizes the orthogonality between the normality of both windows.

Time series analysis has also been used for unsupervised anomaly-based NIDS. These statistical methods are used to predict future values for different network features (SFD or MFD), which are compared with actual, measured values. Records with high prediction errors are flagged as anomalies [54, 164]. Time series have also been proposed to model the likelihood of the alarms generated by an unsupervised Long Short Term Memory Neural Network used as anomaly detector [165]. The method filters the number of alarms, indicating that there is an anomaly when this number deviates highly from the expected value.

We end this description of unsupervised anomaly-based systems with clustering methods. They group data that share similar characteristics with the help of similarity or distance measures [166]. The objective is to identify large or dense clusters, which usually correspond with normal data, flagging as suspicious anything that deviates from those clusters. Those methods assume that there is a minority of attack samples, compared to normal traffic. If this were not the case, expert knowledge would be required to interpret the obtained groups. Depending on the method, thresholds can be applied to density measures [45, 52, 64, 96, 120, 121, 167], flagging observations in low density zones as attacks, or to distances [96, 163, 168], flagging farthest observations (or clusters) from the reference normal points as anomalous.

Supervised approaches have also been proposed for anomaly detection, although rarely. The authors of [135] learn a Random Forest classifier with normal data, using the service identifier (TCP, UDP port number) as the class label. In operation, when a new sample arrives, if the service assigned it by the classifier does not match the specific service to which the sample belongs, an alarm is raised.

Weakly supervised anomaly detectors for NIDS use partially labeled data, aiming to improve the accuracy of unsupervised methods. The authors of [65, 96] apply a one-class SVM classifier using a dataset in which only a few samples are labeled as "normal traffic". In [68], an ensemble of one-class SVMs is used to extract normality models for each network service separately. Proposal [31] also uses an ensemble of one-class SVMs, although it is focused on HTTP attacks. Each model is trained with a different set of features, and the output of all the models is combined to obtain the final classification result. Similarly, the authors of [32] use an ensemble of five Hidden Markov Models (HMM), a probabilistic classifier, in order to predict the probability that a new traffic record is normal. To do that, only payload sequences of normal traffic are used to train the models. Again, their output is combined to obtain a final score. Proposal [169] uses a semi-supervised variant of SOM [157] in which the class information of the labeled records is used to assign labels to clusters. At training time, when a new sample is fed, it is labeled with the class of its nearest cluster (known as the activated cell). Both that and the remaining clusters that have the same class label are then "rewarded", moving their centroids closer to the sample. Semi-supervision is added to deal with unknown clusters (those that are unlabeled), which are always rewarded. The method is also able to merge and split clusters, respectively merging very pure cells (those with a large number of records of the same class) or creating new clusters from impure cells.

The anomaly-based NIDS methods discussed above present some advantages and limitations which, although they are not enough to perform a comparison among them, are worth mentioning. For clustering, the use of feature transformation methods depends on the choice of the distance and the type of clustering, with density methods creating groups of any shape, while distance based methods form spherical clusters. Correlation methods require numerical features for the computation of covariance or correlation matrices but, as their main shortcoming, they lack explainability when attacks are detected, as those matrices are generated from groups of records. Time series analysis demands using numerical data and it is difficult to apply with non-stationary data. However, it is a useful resource to model temporal relations in traffic. Some frequency methods are also able to model temporal relations while working with any kind of data, but they may have poor performance in networks with imbalanced types of normality. Finally, weakly supervised methods require labels and cannot be as easily adaptable to incremental approaches as the other methods.

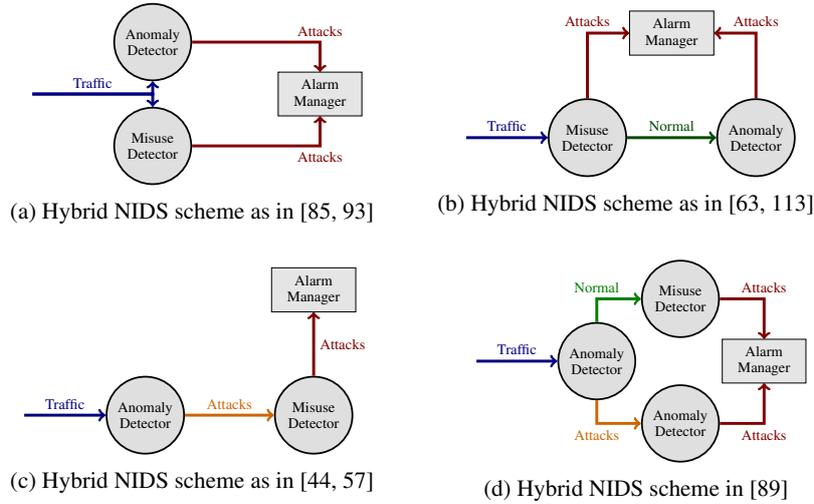

Figure 5: Hybrid-based NIDS schemes

### 7.1.3 Hybrid NIDS

Misuse and anomaly-based methods can be combined into a hybrid detector trying to overcome the limitations of each method separately. For example, to increase detection ability or to keep false alarm rates under control. At the cost of an increased complexity of the NIDS, this combination can be done in different ways, as depicted in Figure 5.

In Figure 5a we can see a simple case in which a misuse-based detector and an anomaly-based detector work in parallel, flagging an observation as malicious if it causes an alarm on any of the two components [93]. The detector described in [85] incorporates both misuse and anomaly approaches in a single data mining method. The system outputs the classification made by the misuse component, and computes the level of deviation from the identified class. Note how this parallel set-up may detect more attacks than a single system, at the cost of an increase in the number of false positives caused by any of the two detectors.

Another way of combining two detectors is concatenating them. The first module of the detector receives traffic records and feeds its output to the second module. Depending on the order in which the misuse and anomaly based detectors are chained, the combination reduces the number of false alarms or increases the detection ability. The architecture proposed in [63, 113] places first the misuse-based system, which receives all network records and identifies well-known malicious behaviors (see Figure 5b). Afterwards, the anomaly detector receives only those records previously classified as normal in order to discover unusual patterns, increasing the detection ability of the NIDS. The inverse approach has been proposed too [44, 57], as depicted in Figure 5c, with the intention of reducing false alarms. The anomaly detector flags suspicious traffic, and afterwards the misuse module confirms and explains the flagged anomalies.

The variant depicted in Figure 5d, and proposed in [89] puts an anomaly detector first. Samples flagged as normal are forwarded to a misuse detector that may identify attacks that passed undetected. Suspicious samples are forwarded to a second, slightly different, anomaly detector, aiming to confirm the attacks and to reduce the number of false alarms. Note that this is the most costly scheme as all input records are analyzed twice.

## 7.2 Batch vs. Incremental Learning

In our taxonomy, the second NIDS classification criterion refers to the different policies used to learn models and keep them updated. Networks, and network traffic, evolve continuously and, therefore, the characteristics of normal traffic and attacks change over time [5]. In this context, the flexibility and adaptability of the data mining algorithms and their tolerance to previously unseen patterns is crucial [27]. However, this fact is rarely mentioned in the literature. In this section we study

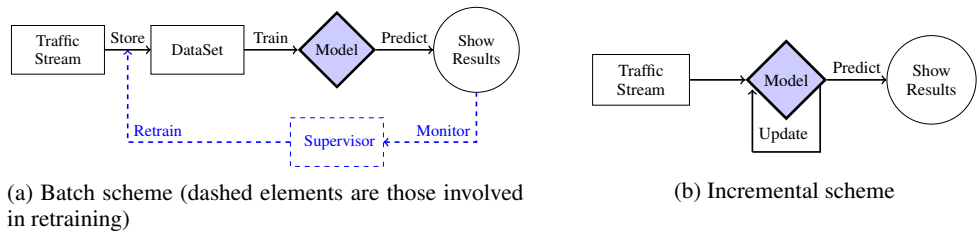

(a) Batch scheme (dashed elements are those involved in retraining)

(b) Incremental scheme

Figure 6: Learning schemes of data mining methods.

the performance and practical usability of two different learning approaches discussed in the NIDS literature: batch and incremental.

### 7.2.1 Batch learning NIDS

Learning in *batch mode* means that the data mining algorithm is applied once to a set of static (training) data, labeled or unlabeled, and after that the results or detectors obtained are used to make predictions for new, incoming data (see Figure 6a). Typically, this type of algorithms require multiple passes over the training data and, since they have not been designed specifically for streaming scenarios, they are usually costly in terms of both time and memory requirements [170]. The effectiveness of these models is high at deployment time, but they do not evolve with time. There is, thus, an implicit assumption of data being generated by a stationary distribution [171]. This is an important drawback, as it seriously challenges the ability to detect previously unseen traffic patterns, that may correspond to new or mutated attacks, or to new kinds of normal traffic.

Most classical data mining methods, independently of their degree of supervision, are based on batch learning. Many NIDS proposals leverage off-the-shelf supervised classification or clustering algorithms and, therefore belong to this category. For example, [86] and [139] rely on well-trained models to perform misuse detection. In a similar fashion, [103] and [128] use a batch approach to model normality in their anomaly detectors. None of these proposals are able to update their detectors automatically.

Other proposals try, instead, to somehow capture and adapt to the traffic changes via retraining [172]. This extension of the batch mechanism adds a supervisor component to the detection framework (see dashed elements of the Figure 6a) that monitors the status of the network or the response of the system, and decides if and when it is necessary to trigger a retraining procedure. Retraining means using the data mining algorithm of choice again, but with a recent batch of stored records, to generate an updated model that replaces the obsolete one. All detectors built using batch learning can be enhanced with retraining. However, this learning mode also has some drawbacks. The selection and storage of the data required to build an updated detector in each retraining phase and, when supervised methods are in use, the labeling of this data (see Section 4.2), are non-trivial tasks that require additional resources [17]. Also, as network traffic changes at non-predictable times, additional complexity in terms of a monitoring mechanism must be put into the NIDS.

Some examples using retraining have been proposed in NIDS. In [45], basic time series analysis is carried out over statistics of grouped flows in time windows (MFD features) to detect changes in flow patterns. Then, for any time window in which changes are detected, a retraining is done over the flows within the time window, using their SFD features. A density-clustering algorithm is applied to rank records by their distance to the largest cluster. Finally, those records farther than a predefined threshold are flagged as anomalies. In [51] a supervisor is added to the proposed NIDS that monitors network alterations such as changes in open ports (available services) or connected devices. It is assumed that those alterations modify the structure of normal traffic, but attacks do not vary. When a change is detected, an auto-encoder is used to learn a new sparse representation of the data reflecting the new traffic characteristics. The same labeled dataset used for the initial training is transformed with the new auto-encoder to build an updated classifier that replaces the previous one.

### 7.2.2 Incremental learning NIDS

Incremental learning algorithms assume that (1) data is unbounded and arrives in a continuous manner [173], and (2) data may be generated by non-stationary processes and, therefore, may evolve over time. Consequently, learning is done continuously with the arrival of new data samples (see Figure 6b). The main difference with retraining strategies is that the model or information stored by the algorithm is updated continuously, instead of being rebuilt from scratch. Adaptation to changes is gradual, which better suits highly variable scenarios such as networks, where stationarity is infrequent [18, 172]. In general, the cost is also lower than that for retraining as less data is incorporated to the model. On the downside, since they learn and provide a response in an online fashion, these algorithms can usually only perform a few passes over the data and they have time and memory limitations [172]. Another disadvantage of incremental learning is convergence time: learning about changes requires some time and, meanwhile, accuracy may be poor. This is particularly true at the initial stages of deployment. Continuous adaptation also makes these systems sensitive to noise and perturbations such as those intentionally generated to evade detectors [174]. As such, incremental approaches require additional evaluation measures that monitor the evolution of the learning process and guarantee its correct behavior before deployment [175].

As the most common approach, we can find a group of unsupervised methods that analyze if each incoming data point in the stream is anomalous, that is, if it is dissimilar to other data points observed previously in the stream. Since they are incremental algorithms, these methods typically only process each data point once, summarize the stream in a representation with small memory requirements that can be easily updated in an online manner, and compare the incoming points with this representation. As the most naive approach within this group, in [162] the authors directly analyze the attribute-value pairs of each new sample and flag data points with infrequent values in certain attributes as anomalies. The stream is thus represented as a table of frequencies, which is updated incrementally with each new sample.

As a slightly more complex approach that follows the same idea, several NIDS papers describe the use of incremental clustering algorithms. These algorithms group the data with similar characteristics and identify attacks as observations that lie far from the existing clusters, or that are located in sparse clusters. Contrary to batch clustering approaches, which group static sets of data into a fixed set of clusters, these methods are designed for streaming environments. As such, the data and cluster structures are represented in different compact manners, i.e., by storing only the objects of the current time window [120], by saving only prototypes or centroids [65, 168], by arranging data into groups of similar data objects (microclusters) represented by a vector of characteristics (mean, variance, etc.) [121, 167], or by using a grid structure in which the data is arranged and summarized using some statistics [52]. Cluster assignments are made online and, additionally, in order to adapt to changes that can happen in the distribution of the data, some algorithms include mechanisms to update cluster assignments by adding and removing clusters as observations arrive. For example, [121] uses a nature-inspired (bird flocking) optimization algorithm to control the evolution of the clusters over time, allowing merge or split operations. In [167] microclusters are represented as Gaussian Mixture Models (GMMs) that can be compared using the Kullback–Leibler divergence and merged if they are sufficiently similar.

A second less common group of unsupervised incremental NIDS uses the temporal information of the stream to identify attacks. Their objective is to incrementally learn and update a model that will somehow describe the normal evolution of the stream. Any observation that does not follow the expected behavior will be flagged as an attack. The difference between the proposals in this group lies mainly in the method used to model the evolution of the stream. Proposal [119] leverages the incremental clustering algorithm of [65] described in the previous paragraph, but analyzes the transitions between cluster assignments of consecutive traffic samples. Attacks are detected when anomalous transitions between clusters occur. Another example is [165], where a Long Short Term Memory (LSTM) network is used to model the evolution of the traffic stream. This is a time series forecasting method, which uses past data to predict future values of the stream. Any observation that is very far from its predicted value is deemed an attack.

Incremental supervised methods have also been proposed in NIDS, although rarely. Again, the main drawback of this kind of methods is the requirement of class labels at operation time, something that is non-trivial. An example of this type of methods is [117], where an online version of the Adaboost algorithm, a classifier based on an ensemble of weak classifier models, is used. The method updates those weak classifiers with the poorest performance each time a new data sample arrives. Other similar examples include [17, 48], which use an ensemble of Very Fast Decision Trees (VFTD), an algorithm that learns classification trees incrementally using small batches of data under the assumption that classes do not evolve with time [176]. In [17], only records with low classification confidence are incorporated to the model in order to capture new classes of traffic.

|  |  | \multicolumn{2}{c}{**Actual class**} |
| --- | --- | --- | --- |
|  |  | **Positive** | **Negative** |
| **Predicted** | **Positive** | True Positive (TP) | False Positive (FP) |
|  | **Negative** | False Negative (FN) | True Negative (TN) |

Table 5: Confusion matrix. Rows correspond to the predicted class and columns to the actual class. The value represented in each position of the matrix indicates the number of instances predicted, as stated by the row, that actually belong to the class indicated by the column.

To overcome the problem of lack of maturity of models in the early stages of incremental learning (when models have not yet been fed with enough data), some off-line data may be provided before deployment. An initial model is trained using batch learning, thus assuming an initial distribution of the data. After that, the model is incrementally updated as new samples arrive [6, 17, 48, 167, 168]. This approach has the advantage of relying on a well-learned detector in the initial phases, while overcoming the limitations of pure batch processing.

# 8 Evaluation Stage

The objective of the evaluation stage is to measure how well the previous stages have performed. Ideally, this evaluation should be as rigorous as possible, using different criteria such as detection accuracy, complexity, adaptability, understand-ability, security, etc.

It is not easy to define valid metrics and comparison criteria for all these characteristics. Let us focus on complexity, very closely related to the ability to deploy a NIDS in a real environment. Only a few papers report complexity computations, and we find them incomplete: computational complexity of the data mining algorithm in use defines only a part of the cost of the system. The remaining steps of the KDD process also have a cost, and implementation issues (such as programming languages used, compilers, libraries, run-time environments etc.) play an important role in the processing ability of the NIDS. A fair evaluation of all these aspects is an extremely complex exercise that requires a comprehensive testing environment –something that is beyond the scope of this survey. Similar limitations could be stated for adaptability, understandability and security.

The predictive performance of a NIDS, however, can be easily measured. Many different metrics have been proposed that provide quantitative values that determine how well malicious behaviours are detected. These metrics can be, and are, easily used as yardsticks to compare different proposals. This is why most authors have mainly paid attention to this performance criterion, and that is why we do the same in this review.

In terms of ML, predictive performance metrics measure how well the predictions provided by an algorithm match with the true labels of the input samples. The confusion matrix, see Table 5, is used to easily visualize the detection performance of a NIDS. For binary detection problems (normal vs. attack), it is common to see the positive class as attacks, while negative means normal activity. For multiclass problems, where more than two classes of traffic can be detected, it is common to use a *one-vs-all* procedure: the positive class is the one that we want to measure, while the negative class groups all the instances of the remaining classes.

As can be seen in Table 6, several metrics (receiving different names) can be derived from this matrix, each one focusing on a different facet of the prediction capacity of a NIDS. Note that some of them, such as Detection Rate (DR), True Negative Rate (TNR), and Accuracy and Precision, must be interpreted as *the higher, the better*, taking values close to 1 when the detector performs very well. Most of these metrics account for the proportion of samples for which the model predicts its class correctly. In contrast, the best detectors should report values close to 0 for False Positive Rate (FPR) and False Negative Rate (FNR), as they represent failure ratios.

| Metric | Formula | Metric | Formula |
|---|---|---|---|
| DR, TPR, Recall, Sensitivity | $\frac{TP}{TP+FN}$ | Precision | $\frac{TP}{TP+FP}$ |
| FAR, FPR, Specificity | $\frac{FP}{FP+TN}$ | Accuracy | $\frac{TP+TN}{TP+TN+FP+FN}$ |
| TNR | $\frac{TN}{FP+TN} = 1-FPR$ | Error Rate | $\frac{FP+FN}{TP+TN+FP+FN} = 1-Acc$ |
| FNR, Miss Rate | $\frac{FN}{TP+FN} = 1-TPR$ | F-score/F-measure | $2 * \frac{Precision * Recall}{Precision + Recall}$ |

Table 6: Common evaluation metrics derived from the confusion matrix

Another common metric is the ROC curve. It visually shows the trade-off between the TPR and FPR of classifiers, by means of a bi-dimensional plot, for different values of the decision threshold. The area under the ROC curve (AUC) is also used as a metric to evaluate NIDS.

Not all the metrics used by the machine learning community are meaningful for the specific NIDS problem. Network traffic is a highly unbalanced scenario in which normal traffic is much more common than attacks. Moreover, cyberattacks may have a severe impact on the victim and, thus, metrics used for NIDS should reflect the effects of not detecting an attack (false negatives). Accordingly, the use of some metrics such as accuracy, which may not properly reflect the performance in imbalanced scenarios [177], is not recommended. Nevertheless, several NIDS articles report it [44, 57, 91, 125, 134, 154].

The most common metrics reported in the surveyed NIDS literature are DR (Detection Rate) and FPR (False Positive Rate), as they inform about complementary detection abilities. DR measures how well the system identifies abnormal behaviours as attacks, while FPR measures the ratio of normal observations incorrectly flagged as attacks. The goal of a good IDS is to maximize the former while minimizing the latter. Summary metrics that combine others (such as the F-measure) are also used, but should be considered as supplementary information due to its poor explainability. ROC curves are useful as they provide a graphical way to compare detectors [52, 167]. However, when plots are very similar, additional metrics may be needed for further clarification.

## 9 Discussion, Conclusions and Open Issues

An ideal NIDS should fulfill a large set of desirable characteristics including, among others, the following: early detection of attacks, high attack-detection coverage (ability to detect the maximum number of attacks), high detection rate with low false alarm rate, ability to scale according to network requirements, and robustness to attacks targeting the detector. In this section we summarize the conclusions extracted from our literature review, as well as a collection of open issues and challenges identified, which may constitute future lines of research in the area of network security.

In order to have a general view of the use by NIDS authors of the techniques discussed in this paper, we have rendered two graphs (see Figures 7a and 7b) that represent the paths followed when designing (and evaluating) a NIDS, for both misuse and anomaly-based systems. In these graphs, each level corresponds to a different phase of the KDD process, and the nodes within each phase correspond to the methods used in the surveyed articles. Nodes are joined by arrows, whose thickness represents the proportion of works that apply a particular (destination) method after the (source) method of the previous phase. To allow a better understanding of the graphs, we have simplified them by removing nodes for methods with low utilization (below 5%). In general, main paths (sequences of thick arrows) are characterized by the presence of acronym "ND" (method Not Discussed for that phase), due to two main reasons: (1) many articles do not mention the methods applied to the data, and (2) oftentimes, data mining algorithms are applied directly to "cooked" records obtained from public datasets.

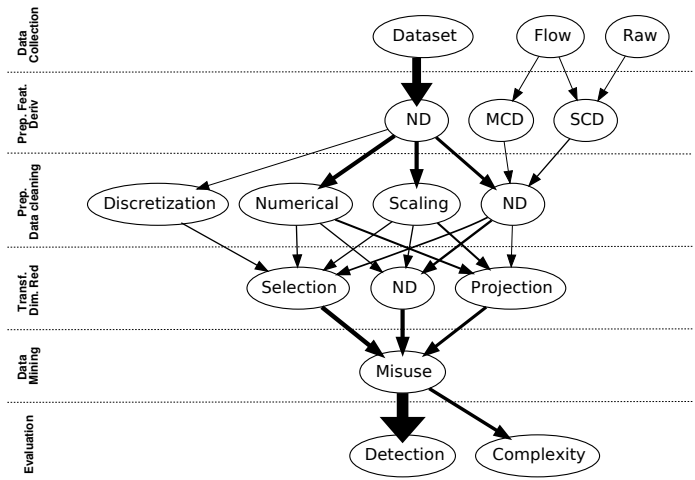

(a) misuse approaches

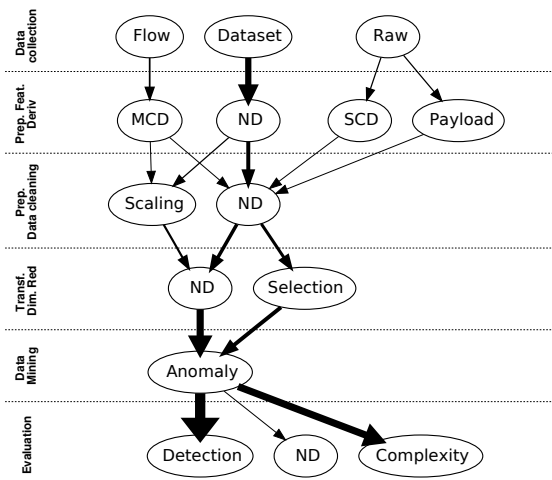

(b) anomaly approaches

Figure 7: Paths along KDD of misuse and anomaly surveyed approaches. Levels represent the different phases of KDD. Nodes represent methods used at each stage. Edge thickness depicts the proportion of the surveyed works using the method below.

## 9.1 Reproducible research

A first conclusion of our literature review that can be inferred from the previous paragraph is that most papers provide insufficient detail about their pipeline and, more specifically, about the procedures used for feature derivation, transformation and validation. In some cases, this is due to the use of public datasets in which all this work has already been done. However, we have found several articles that apply data mining algorithms that can only deal with numerical data over the set of heterogeneous records provided by a public dataset, without specifying any feature removal or feature transformation method. In some other cases, transformations are vaguely described: "we transformed categorical variables to a numerical representation", without stating how. Such a lack of details about the procedures carried out to build NIDSs hinder their analysis, evaluation and compar-ison. In this sense, as a recommendation for future work in the area, more effort is required towards favouring reproducible research, especially by providing all the details of the whole procedure used to build a NIDS.

### 9.2 Data obtention, feature extraction and public datasets

Graphs 7a and 7b show how, in the first phase of KDD (data collection), most NIDSs use public datasets for both model construction and evaluation purposes. In Section 4.3, we have already explained how these datasets were generated, also pointing out their deficiencies (short capture periods, discontinued captures, uncertainty during the labeling process and lack of realism, among others). A NIDS that works successfully with any of these datasets is not guaranteed to operate equally well in real scenarios. Thus, it is not possible to perform a proper evaluation of NIDS with a single dataset; the use of various datasets, with different characteristics, has to be considered. Ideally, datasets which satisfy most of the mentioned characteristics, and that reflect the current use of the network by applications/nodes, are desirable. These databases should be captured at different network locations, and should thus provide different views of the network.

Another potential weakness related to the use of public datasets lies in that each one provides a particular set of already derived features, and most researchers work only with those. Available features may be useless when trying to detect newer attacks, or may contain data that is impossible to obtain in real scenarios, for example labels provided by human experts. To guarantee the derivation of custom features and provide more flexibility to researchers, the availability of raw packet data in public datasets is a must. Indeed, a few NIDS authors derive their own set of features from raw data (see Figures 7a and 7b), normally SFD, MFD or payload features, that have shown to improve the accuracy of the detector [50].

Regarding payload features, we have already mentioned that the trend is that most applications interchange data in a secure, encrypted form. Nonetheless, most public datasets do not contain this kind of traffic (another reason that makes them unrealistic). This absence of encryption has allowed the extraction of payload features, useful to detect some attacks that leave their footprint in the content of a connection. However, either ciphered or in cleartext, obtaining these features is barely possible in real world high-speed networking scenarios. Efficient capture and decryption mechanisms are required to extract the payload of plaintext connections with low overheads. A step towards a solution may be the use of probes installed at the hosts (that is, once the data is received and decrypted), but at the cost of increasing network traffic (payload still has to be sent to the NIDS for analysis) and losing some early detection capabilities.

Precisely because dealing with payload data is anything but easy, many authors have limited their detectors to use SFD and MFD features in order to identify point and collective malicious patterns. These derived features are usually very simple: flow, packet or byte counts and average rates. We think that more sophisticated variables could be useful to increase detection performance, widen the spectrum of identified attacks, harden the security of a NIDS and identify evasion attempts. For example, features that summarize the additional traffic activity caused by a network connection (number of different service requests, such as DNS, ARP, etc.), together with indicators of causal relationships, may also be useful for the detection of contextual patterns of cyberattacks. In general, the extraction and use of additional SFD and MFD features needs to be further explored, when dealing with different kinds of attacks and evasion scenarios [178].

### 9.3 Feature transformation and dimensionality reduction

Our graphs (see Figures 7a and 7b again) show how feature scaling methods or transformations of categorical features into numerical representations are more common in misuse systems than in anomaly detectors. The main reason is that some data mining algorithms require datasets which are free of outliers, balanced in terms of labels and homogeneous in terms of data type natures (numerical or categorical) to work properly.

Numerical transformations of categorical data allow the use of data mining algorithms that can only deal with numerical data. Some of these representations, such as dummy features or frequency encoding, substantially increase the length of the feature vectors. Consequently, dimensionality reduction methods, such as FSS and projections, become mandatory. Some anomaly-based NIDS performs this reduction through the manual removal of unwanted features.

This combination of techniques is used to enable the use of data mining algorithms and also to increase their detection accuracy. While this may be a good solution for static environments, what happens when network traffic changes? New threats may leave traces in features incorrectly modified (discretized or scaled, with unknown categories...) or selected for removal, as this was done using criteria based on obsolete traffic patterns. As a solution to this problem, dynamic preprocessing and transformation techniques need to be explored.

### 9.4 Data mining algorithms and learning schemes

As regards the data mining phase, misuse systems are mainly based on supervised classifiers, while unsupervised and weakly supervised methods are the choice for anomaly detection. Based on our review of the literature, we have seen that many works based on supervised classifiers have erroneously used the term "anomaly detection" to refer to their detection approaches. In view of this, we have concluded that the existing definitions for the concepts of misuse-based and anomaly-based NIDS do not allow a clear categorization of the existing NIDS, and in this work we have provided a new taxonomy to avoid such miscategorizations.

Also, we have pointed out the limitations of batch learning methods when dealing with evolving traffic patterns. Despite this, we have shown that most published works use batch learning, implicitly assuming unrealistic, static traffic patterns. Sometimes retraining mechanisms are implemented and the training of the new model is done while the NIDS is operating with the obsolete model, which in fact is dangerous as the number of false alarms may increase, while attacks may pass undetected. The number of incremental learning methods considering networks as continuously evolving scenarios is very small for both misuse and anomaly detectors. For misuse incremental detectors, human intervention is often required to label new or unknown records before updating the model. Anomaly-based incremental detectors are prone to generating many false alarms and may be easily evaded. In general, we think that incremental learning approaches for NIDS, taking into account their risks and benefits, should be investigated further.

The small interest paid by the authors to temporal dependencies in the input data when designing detectors also needs to be highlighted. Indeed, we have briefly discussed this issue as a missing aspect in the feature derivation step (see Section 9.2 for this discussion). A few works apply elementary time series analysis to exploit temporal relations, but a thorough study of the temporal relations is missing. Most proposals analyze traffic data once connection statistics have been computed, i.e., detection is carried out over connections that have already occurred; therefore, an attack is detected after it starts taking place. Performing early detection of malicious traffic with models that exploit temporal patterns and dependencies among different communication protocols could be of interest since it can limit attack consequences. For example, the analysis of ARP flows could be used to detect some man-in-the-middle attacks at an early stage. In general, we think that these approaches require further attention.

### 9.5 NIDS evaluation beyond accuracy

A common issue when comparing NIDS papers is the lack of an adequate framework to evaluate and compare different proposals. Authors often use unfit metrics (we have already discussed how some metrics that are commonplace in ML literature are meaningless in NIDS) and obviate the evaluation of important properties beyond those related to accuracy. Complexity, adaptation ability and resistance to attacks towards the NIDS itself should be assessed.

Complexity is sometimes discussed in the NIDS literature (see the lower part of Figures 7a and 7b) but, in our opinion, this issue deserves further research efforts. A theoretically excellent detector is not practical if response times are high and attacks are reported only after the attacker has fulfilled his/her purpose. An effort to collect the *training* complexity of NIDS proposals was made in [12], but the computational cost associated to this phase is only critical for retraining procedures. In addition, and as we have seen throughout this survey, the operational cost of NIDS detectors may be dependent of any other steps apart from detection (data preprocessing, data transformation...). Such costs should be reported, as they would indicate whether or not a NIDS may work in real time, and with how many resources.

The evaluation of the adaptability of incremental learning approaches is also an open issue [179]. These learning mechanisms need to be robust to noise (or attacks) while being able to adapt to newer traffic patterns. In other words, a balance between sensitivity and robustness has to be found by means of a proper evaluation of different variability scenarios, a fact barely discussed in NIDS that use incremental learning.

Finally, assessments of the security features of NIDS proposals are also infrequent. As can be seen in Figures 7a and 7b, there is no node representing this aspect of the evaluation. A NIDS plays an important role in the protection of the corporate network, and should be resistant against attacks that target it [146]. To the best of our knowledge, the first article discussing this risk dates from 2006 [180]; posterior research also tackles this issue [50, 181–183]. After that, a reduced number of NIDS proposals incorporate some protection mechanism, but of limited scope or without performing any evaluation [32, 48, 54]. More effort must be placed on researching NIDS security, as well as on applying a framework to evaluate it [184].

## Acknowledgments

This work has received support from the TIN2016-78365-R (Spanish Ministry of Economy, Industry and Competitiveness http://www.mineco.gob.es/portal/site/mineco) and IT-1244-19 (Basque Government) programs. Borja Molina Coronado holds a predoctoral grant (ref. PRE_2019_2_0022) by the Basque Government.